\documentclass{pnastwo}

\usepackage[dvips]{graphicx}
\usepackage{amssymb,amsfonts,amsmath,bm}

\def\Real{\mathrm{Re}_\alpha}
\def\Realon{\mathrm{Re}_\alpha^{\rm on}}
\def\Relam{\mathrm{Re}_\lambda}
\def\vs{v_{\rm s}}
\def\vn{v_{\rm n}}
\def\vels{{\bf v}_{\rm s}}
\def\veln{{\bf v}_{\rm n}}
\def\velL{{\bf v}_{\rm L}}
\def\rhos{\rho_{\rm s}}
\def\rhon{\rho_{\rm n}}
\def\fo#1{{\bf F}_{\rm #1}}
\def\unis{\hat{\bf s}}
\def\oms{\bm{\omega}}
\def\unioms{\hat{\bm{\omega}}}
\def\Om{\bm{\Omega}}
\def\rvec{{\bf r}}
\def\Oms{\Omega_{\rm s}}
\def\Omi{\Omega_{\rm i}}
\def\Omf{\Omega_{\rm f}}
\def\Vf{V_{\rm f}}
\def\Ninj{{\cal N}_{\rm i}}
\def\aam{\alpha_{\rm am}}
\def\aen{\alpha_{\rm en}}
\def\taam{\tilde\alpha_{\rm am}}
\def\taen{\tilde\alpha_{\rm en}}
\def\Cam{C_{\rm am}}
\def\Cen{C_{\rm en}}

\url{www.pnas.org/cgi/doi/10.1073/pnas.0709640104}
\copyrightyear{2008}
\issuedate{Issue Date}
\volume{Volume}
\issuenumber{Issue Number}
\begin{document}

\title{Quantum turbulence in superfluids with wall-clamped normal
  component}

\author{Vladimir Eltsov\affil{1}{O.V. Lounasmaa Laboratory, Aalto
    University, P.O. Box 15100, FI-00076 AALTO, Finland},
Risto H\"anninen\affil{1}{},
\and
Matti Krusius\affil{1}{}}

\contributor{Submitted to Proceedings of the National Academy of Sciences
of the United States of America}

\maketitle

\begin{article}
\begin{abstract}
  \boldmath In Fermi superfluids, like superfluid $^3$He, the viscous
  normal component can be considered to be stationary with respect to the
  container. The normal component interacts with the superfluid component
  via mutual friction which damps the motion of quantized
  vortex lines and eventually couples the superfluid component to the
  container. With decreasing temperature and mutual friction 
  the
  internal dynamics of the superfluid component becomes more important compared to the damping
  and coupling effects from the normal component. This causes profound
  changes in superfluid dynamics: the temperature-dependent transition
  from laminar to turbulent vortex motion and the decoupling from the
  reference frame of the container at even lower temperatures.
\end{abstract}


\abbreviations{NMR, nuclear magnetic resonance; CF, counterflow}

\dropcap{I}n this paper we consider the motion of quantized vortices in superfluids, where the normal component
is clamped to the walls of the container, that is, it is stationary in a
reference frame moving with the wall. This situation can be experimentally
realized in superfluid $^3$He. In such systems turbulent motion can
occur only in the superfluid component and thus, in principle, is easier to
analyze. Here the role of the normal fluid is twofold: First, it provides friction in the
superfluid motion, which is mediated by quantized vortices and works over a
wide range of length scales. Second, it provides a coupling to the container walls,
which acts uniformly over the whole volume of the superfluid. This is quite
unlike classical turbulence, where viscous dissipation operates only at the
small Kolmogorov scale and coupling to the walls is provided by thin
boundary layers.

%In the B phase of superfluid $^3$He mutual friction decreases by about six
%orders in magnitude from the superfluid transition temperature down to the
%lowest laboratory temperatures of today. 
As a result, a variety of new
phenomena is observed in experiments and numerical simulations as a
function of temperature. These are controlled by the normal-fluid density.
At the highest temperatures turbulent motion is suppressed completely. When
the temperature decreases, a sharp transition to turbulence is seen. The
transition can be characterized by a superfluid Reynolds number, which is
composed of the internal friction parameters of the superfluid and is
independent of velocity. When turbulence is triggered by a localized
perturbation of the laminar flow, the critical value of the Reynolds number
is found to scale with the strength of the perturbation.

When the temperature decreases further, friction from the normal
component rapidly vanishes. The overall dissipation rate in quantum
turbulence remains nevertheless finite, owing to the contribution from the
turbulent energy cascade, and reaches a temperature-independent value in the zero-temperature
limit. This zero-temperature dissipation can be characterized by an effective viscosity or
friction. It is found, however, that coupling to the walls is not
essentially improved by the turbulence and can potentially become very small. At the lowest
temperatures the concept of a single effective friction breaks down;
for the proper description of quantum-turbulent flows one then has to introduce a
separate effective friction for momentum exchange between the
superfluid and the boundaries.

\section{Coarse-grained superfluid dynamics and mutual friction}

According to the two-fluid model, the flow in superfluids at
finite temperatures involves separate motions of the normal and superfluid
components. These components have densities $\rhon$ and $\rhos$ and
velocities $\veln$ and $\vels$, respectively. In this paper we are concerned only with
processes which happen at nearly constant temperature. In such simplified
cases the equations of the two-fluid hydrodynamics can be understood
to originate from the Euler equation for the superfluid component and from the
Navier-Stokes equation for the normal component, with the mutual friction force providing the
interaction between the two components \cite{soninRMP}.

\begin{figure}
\centerline{\includegraphics[width=\linewidth]{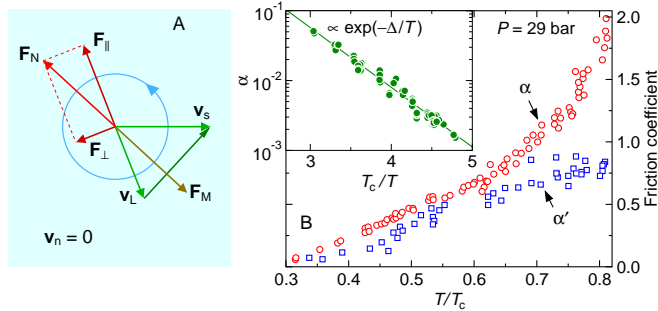}}
\caption{Mutual friction in $^3$He-B. ({\it A}) Forces acting on a vortex
  moving with respect to the superfluid and normal components: Magnus force
  $\fo{M}$ and reaction of the normal component $\fo{N}$, with $\fo{M} = - \fo{N}$. ({\it B})
  Measured values of the mutual friction coefficients as a function of
  temperature. The main panel shows $\alpha$ and $\alpha'$ at $T > 0.3\,T_{\rm
    c}$ as determined from the damping of an oscillating diaphragm \cite{bevanJLTP}. The \textit{insert}
  shows values of $\alpha$ at $T < 0.3\,T_{\rm c}$ as derived from the decay of
  laminar spin-down from rotation \cite{LaminarDecay}. The line is a fit to the theoretical
  dependence $\alpha = \alpha_0 \exp(-\Delta/T)$ with $\alpha_0 = 21$ as
  fitting parameter and the bulk energy gap $\Delta = 1.968 T_{\rm c}$
  at 29\,bar pressure.
\label{friction}}
\end{figure}

The mutual friction interaction is mediated by quantized vortex lines. When a straight
vortex moves with velocity $\velL$ in a superfluid it experiences two
forces (Fig.~\ref{friction}A): The first one is the Magnus lift force from the
superfluid component $\fo{M} = \kappa \rhos (\vels - \velL) \times
\unis$. Here $\kappa$ is the quantum of circulation and $\unis$ is a unit
vector along the vortex core. The second force arises
from the scattering of the quasiparticles, which form the normal component,
from the vortex cores. It has the components $\fo{N} = \fo{\|} + \fo{\perp}$, where $\fo{\|} \propto
(\veln-\velL)_\perp$ and $\fo{\perp} \propto \unis\times
(\veln-\velL)$ (the subscript '$\perp$' denotes projections to the
plane perpendicular to the vortex core). Vortex mass can usually be ignored and the equation of
motion reduces simply to the force balance, $\fo{M} + \fo{N} = 0$. Solving this for
$\velL$ yields
\begin{equation}
\velL = \vels + \alpha'(\veln - \vels)_\perp + \alpha \unis \times (\veln -
\vels).
\label{eq:velL}
\end{equation}
Here $\alpha(T,P)$ and $\alpha'(T,P)$ are the mutual friction parameters,
characteristics of the superfluid, which describe the interaction of thermal
quasiparticles with the vortex cores. Inserting $\velL$ from
Eq.~\eqref{eq:velL} to the expression for $\fo{M}$ we find that in the
mutual friction force $\fo{N}
= -\fo{M}$ the term with $\alpha$ is always directed opposite to the so-called counterflow velocity $\vels -
\veln$ and thus leads to dissipation, while the term with $\alpha'$ is
non-dissipative.

The mutual friction force can be averaged over vortex lines if they are
locally parallel to each other. In this case we get a force per unit mass
of the superfluid component
\begin{equation}
\fo{ns} = \alpha \unioms \times[\oms \times (\vels - \veln)]
 -\alpha' (\vels - \veln) \times \oms~.
\label{eq:fns}
\end{equation}
Here the vorticity $\oms = \nabla \times \vels$ and $\unioms$ is a unit vector
in the direction of $\oms$. Note that the friction coefficients in
Eq.~\eqref{eq:fns} are the same as in Eq.~\eqref{eq:velL} only because
complete local polarization of vortex lines is assumed. Later we will
introduce effective mutual friction parameters to describe
phenomenologically situations where this is not the case.

If the vortex line is curved, then its tension $T_{\rm v} = (\kappa^2
\rhos/4\pi) \ln(\ell/a)$, originating from the quantized superflow around the
core, leads to an
additional force acting perpendicular to the core. Balancing it against
the Magnus force and averaging over the vortex lines we get the contribution
\begin{equation}
\fo{tens} = -\lambda \oms \times (\nabla \times \unioms)~,
\label{eq:tens}
\end{equation}
where $\lambda = (\kappa/4\pi) \ln(\ell/a)$, $\ell$ is the intervortex
distance and $a$ is the core size. Including the tension force to the balance
of forces acting on a vortex modifies also the expression for the mutual
friction force, but we will not consider this here. Since $\fo{tens}$ has the
small prefactor $\lambda \sim \kappa$, in many cases it can be
neglected. However, there are situations where it is essential, like in the
Kelvin-wave instability \cite{Glaberson} or in the angular momentum balance of the
propagating vortex front \cite{FrontNatComm}, considered below.

Inserting the $\fo{ns}$ and $\fo{tens}$ forces to the Euler equation, we get the
coarse-grained hydrodynamic equation for the superfluid component
\begin{equation}
\frac{\partial \vels}{\partial t} + \nabla(\mu + \vs^2/2) =
\vels \times \oms + \fo{ns} + \fo{tens}~,
\label{eq:vs}
\end{equation}
where $\mu$ is the chemical potential. To account for the normal component, one inserts the
$\fo{ns}$ force with a negative sign to the Navier-Stokes equation,
where the viscous term then damps the drive from mutual friction. By
comparing the magnitudes of these two terms, it is possible to work out a
criterion, when the normal component can be
considered not to respond to the motion of the
superfluid component in the hydrodynamic regime \cite{rpp}: $\nu / \kappa \gg (\rhos/\rhon)
\alpha$. Here $\nu$ is the kinematic viscosity of the normal
component. The Fermi superfluid $^3$He has a large value of $\nu / \kappa
\sim 10^3$, which satisfies this criterion and allows us to consider the
normal component as stationary. This is quite unlike Bose superfluids where usually $\nu \sim \kappa$ \cite{he4quant}. At low temperatures in the ballistic regime of quasiparticle motion one might expect that the drag from
the quasiparticle gas on the vortex lines depends strongly on the ratio of
the intervortex distance to the scattering cross-section in the quasiparticle-vortex scattering process. It turns out that at the typical experimental values of $\ell \sim 0.1\,$mm this consideration is not important. Numerical calculations at higher
vortex densities show, however, some non-trivial effects in quasiparticle
motion \cite{sergeev}.

Values of the mutual friction coefficients $\alpha$ and $\alpha'$ are thus
important for understanding the dynamic properties of the superfluid. In the isotropic B phase of superfluid $^3$He they have
been measured in wide temperature and pressure ranges by the
Manchester group \cite{bevanPRL,bevanJLTP} (Fig.~\ref{friction}B). The results are in fair agreement
with the theory of mutual friction which includes the Iordanskii force
from the scattering of bulk thermal quasiparticles by the flow field around
the vortex \cite{Iordanskii} and the Kopnin force originating from the spectral flow of the
vortex-core-bound fermions and their scattering from the bulk
quasiparticles \cite{kopnin-rpp}.
The energy gap $\Delta$ in the spectrum of bulk quasiparticles is the
main source for the fast variation of mutual friction with temperature.
The original measurements left an uncertainty \cite{bevanJLTP} whether the accepted value of the
gap should be renormalized to fit the experimental
data. New measurements of friction to lower temperatures
\cite{LaminarDecay,AndreevPRB2} leave no
doubt that the bulk $\Delta$ properly describes the temperature
dependence, in accordance with theory (insert in
Fig.~\ref{friction}B).

An unsolved problem, both experimentally and theoretically, is the behavior
of mutual friction at very low temperatures, below about $0.15T_{\rm
  c}$. There is a prediction of a new friction mechanism \cite{silaev}, which depends on
the acceleration of vortex lines, but not on temperature and thus
allows for a finite dissipation even at zero temperature through the emission of
non-thermal quasiparticles from the vortex cores. There are
observations of temperature-independent contributions to friction \cite{AndreevPRB2,FrontNatComm}, but
it is not yet clear, whether they can be attributed to a mechanism of this kind.

\section{Transition from laminar to turbulent dynamics}

\subsection{Criterion for turbulence}
To establish the ranges of stability for laminar and turbulent flows in
superfluids with a stationary normal component, one can apply the same dimensional arguments as used
in classical hydrodynamics, by introducing the Reynolds number Re. The classical Reynolds number ${\rm Re} = UR/\nu$ has the physical meaning of being the ratio of the magnitudes of the inertial and dissipative
terms in the Navier-Stokes equation. Here $U$ is the typical velocity of
the flow and $R$ is the spatial scale. Turbulent motion is observed for
${\rm Re} \gg 1$. In superfluids, starting from Eq.~\eqref{eq:vs} for
the superfluid velocity, we put $\veln = 0$ and omit the tension term so that we get the equation
\begin{equation}
\frac{\partial \vels}{\partial t} + \nabla(\mu + \vs^2/2) =
(1-\alpha')\vels \times \oms +
\alpha \unioms \times(\oms \times \vels)~.
\label{eq:vs0}
\end{equation}
Here the inertial term is the same as in the Navier-Stokes equation, but
renormalized with the prefactor $1-\alpha'$. The dissipative term, proportional
to $\alpha$, has a completely different structure. In fact, both the
inertial and dissipative terms have the same scaling $\sim U^2/R$ and thus
their ratio becomes an internal parameter of the superfluid, independent
of the velocity and the geometry of the flow \cite{FinneNature}
\begin{equation*}
\Real = \frac{1-\alpha'}{\alpha}~.
\label{eq:real}
\end{equation*}
This parameter has the same physical meaning as the Reynolds number in
classical hydrodynamics and thus we call it the superfluid Reynolds
number. In analogy to classical turbulence we can expect superfluid flow to
be turbulent when $\Real \gg 1$ and laminar in the opposite limit. Measurements show, however, that the value of
$\Real$ at the transition from laminar to turbulent dynamics, which
we call the onset value $\Realon$, is actually for many types of
flow much closer to unity than in classical systems. Thus we write the criterion for
superfluid turbulence as
\begin{equation}
\Real \gtrsim 1~.
\label{eq:crit}
\end{equation}
It is possible to approach the limits for quantum turbulence
also from a microscopic point of view. The essential processes are vortex reconnections, growth of vortex rings, and Kelvin
waves on vortex lines. One can estimate within some model when these
processes proliferate or when they are suppressed by mutual friction.
Different models have been considered: the pseudo-Vinen equation
\cite{kopnin-crit,rpp}, damping of Kelvin waves on vortex bundles
\cite{henderson}, and the reorientation of a vortex ring in circulating flow
\cite{mult-prl}. In all cases the same criterion Eq.~\eqref{eq:crit} is
obtained.

\begin{figure}
\centerline{\includegraphics[width=\linewidth]{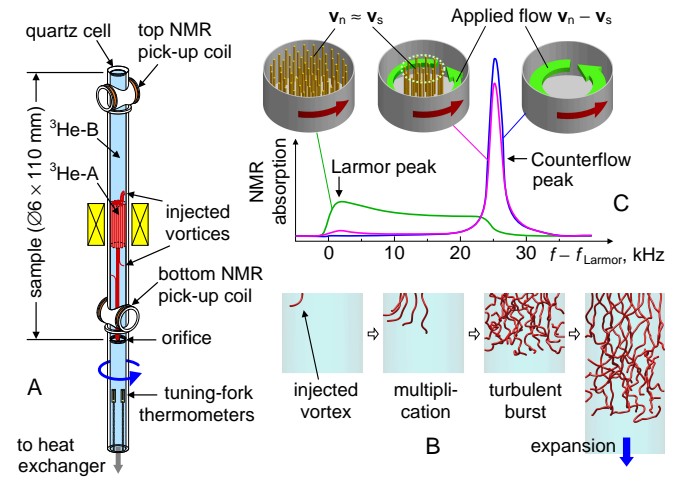}}
\caption{Experimental setup and principles. (\textit{A})
  The sample is contained in a smooth-walled quartz cylinder, separated from the
  volume in contact with the rough surfaces of a sintered heat exchanger by a plate
  with a small orifice. The temperature is determined from the resonance width
  of a quartz tuning fork oscillator. The A phase of superfluid $^3$He
  can be stabilized in the central section of the sample with a magnetic
  field. The upper B-phase section is about 1\,cm shorter than the lower. In rotation the AB interface becomes unstable at a well-defined velocity and a bundle of $\sim 10$ closely packed vortex loops is
  injected from the A into the B phase, as shown for the upper AB interface.
  In the lower B-phase section an alternative way is illustrated for putting vortices
  in applied flow. It uses slowly evolving remanent vortices which terminate
  at the cylinder wall. (\textit{B}) Turbulent evolution of the injected vortex loop(s). (\textit{C}) The number of
  vortices within the NMR pick-up coil is determined from the height of
  the so-called counterflow peak in the NMR spectrum of $^3$He-B \cite{rpp}.
\label{expsetup}}
\end{figure}

When the vortex tension is included in Eq.~\eqref{eq:vs0}, two additional
dimensionless parameters can be defined. The ratio of the inertial term to
the tension term is
\begin{equation*}
\Relam = UR/\lambda~.
\label{eq:relam}
\end{equation*}
An alternative, but similar definition of ${\rm Re}_{\rm s} = UR/\kappa$ is
often used. Owing to the functional similarity of this expression to the
classical Reynolds number, this combination is also called a superfluid
Reynolds number. (To add to the confusion, in superfluids with a mobile normal
component even more 'superfluid Reynolds numbers' can be defined.) The
physical meaning of $\Relam$ is different, however. The condition $\Relam \sim
1$ coincides with the Feynman criterion which specifies when the first vortex line becomes
stable in the flow \cite{rpp}. When $\Relam \gg 1$ the flow can support many vortex
lines. This is a necessary condition for the applicability of the
coarse-grained hydrodynamic equations. Here we consider only cases
where this condition is fulfilled.

\begin{figure}
\centerline{\includegraphics[width=\linewidth]{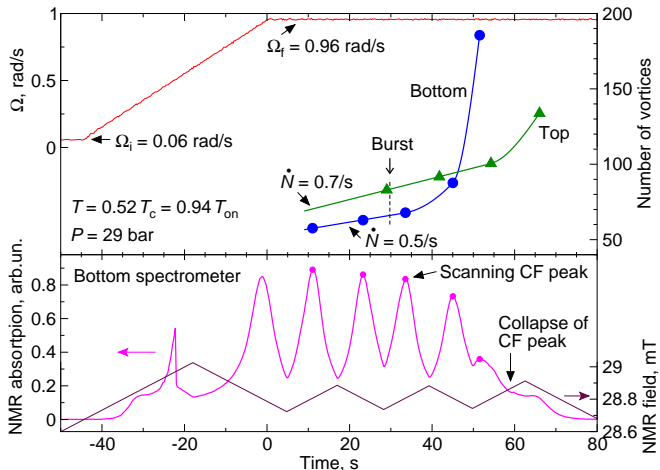}}
\caption{Evolution to a turbulent burst. This measurement starts from the equilibrium vortex state at $\Omi
  = 0.06\,$rad/s, where some vortices connect to the cylindrical wall, owing to the residual inclination (of $\approx 1^\circ$) of the long cylinder with respect to the rotation axis. When $\Omega$ is rapidly ramped to $\Omf$, the curved vortex ends become unstable in the applied flow and generate new vortices, first in slow
  single-vortex processes, and finally in a turbulent burst, when the newly created
  vortices interact turbulently. The number of vortices within the top
  and bottom NMR coils (top panel) are obtained from the height of the counterflow (CF) peak,
  which is continuously scanned, as shown in the bottom panel. In
  laminar flow the height of the CF peak does not decrease with
  time, while here it collapses suddenly.  In this measurement there is not A-phase barrier layer; thus
  from the arrival times of the vortex fronts to the NMR coils one obtains the moment when the turbulent burst happened (marked
  with an arrow in the upper panel) and the location of the burst, which in
  this measurement happens to be 4\,cm above the bottom NMR coil.
\label{slowformation}}
\end{figure}

Finally, the ratio of the tension term to the dissipation term $\Real/\Relam$
is the parameter which controls the superfluid decoupling phenomenon \cite{decouplingPRL}, as
discussed later in this paper.

\subsection{Measurement of the transition to turbulence in rotating flow}
Many conventional methods for generating turbulence in classical
liquids or in superfluid $^4$He are not applicable in $^3$He-B owing to the special requirements at
ultra-low temperatures and the large viscosity of the normal
component. The most detailed measurements of the transition to turbulence have been performed in rotating flow with a cylindrical sample container of radius $R$ and oriented along the rotation axis, Fig.~\ref{expsetup}A. The walls of the
container are sufficiently smooth to avoid vortex pinning and surface friction. This is possible with $^3$He-B since the vortex core radius is relatively large, $a\sim 0.1\div 0.01\,\mu$m, depending on pressure.

\begin{figure}
\centerline{\includegraphics[width=\linewidth]{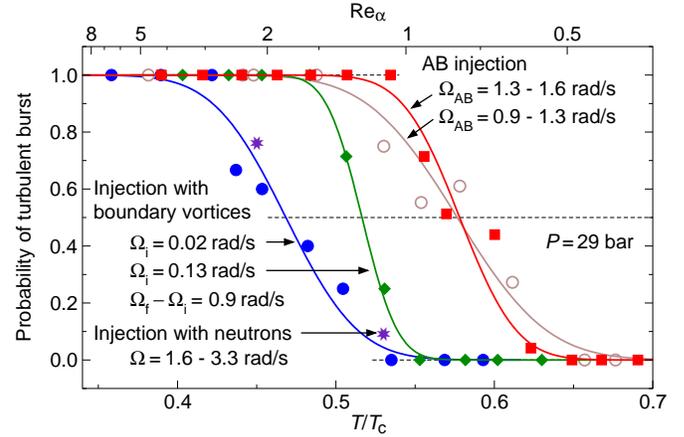}}
\caption{Transition to turbulence with decreasing mutual friction
  damping. The different symbols represent results from different types of vortex-injection measurements.
  The probability to start turbulence is
  determined from about 5 measurements for each set of conditions. The solid
  curves are fitted cumulative normal distribution functions from
  which $T_{\rm on}$ and $\Realon = \Real(T_{\rm on})$ are determined.
  \label{probabs}}
\end{figure}

In uniform rotation the normal component of $^3$He-B is in solid-body rotation: $\veln = \Om \times \rvec$. The most obvious laminar solution of Eq.~\eqref{eq:vs} is then $\vels = 0$. The formation of
quantized vortices 'out of nothing' at moderate flow velocities $\vn
\lesssim 1\,$cm/s is prohibited by a large energy barrier and thus this metastable
vortex-free state can persist for the duration of the experiment in the
absence of seed vortices \cite{Ruutu-vortform}. Another state of laminar flow is solid-body rotation of the superfluid component at an angular velocity $\Oms$: $\vels = (\Oms\hat{\bm{\Omega}})\times\rvec$. In this state the
sample is filled with rectilinear vortices at a density
$2\Oms/\kappa$, aligned along the rotation axis. If the rotation drive
$\Omega$ is increased from $\Oms$, then the vortices move towards the axis
or if $\Omega$ is reduced they move away from it. In both cases the
velocity of the laminar expansion or contraction of such a cluster of
parallel vortex lines is determined by $\alpha$ \cite{LaminarDecay}.

One more example of laminar flow becomes important in the laminar regime of vortex motion $(\Real < 1)$, when short seed vortices initially occupy only part of the height of the cylinder, while the rest remains
vortex-free. These vortices bend to the cylindrical wall, as seen in Fig.~\ref{expsetup}A. They continuously expand, until they become fully rectilinear line vortices, as their ends move with an axial velocity $V_{\rm lam} \approx \alpha \Omega R$ (or
somewhat smaller, depending on the curvature of the vortex) \cite{precess-jltp,karimaki}.

Among the different types of turbulence in rotating flows here we discuss only two,
Fig.~\ref{expsetup}B. The first is the turbulent burst \cite{Finne-turbmeas-jltp,rpp}, when a few closely
packed vortex loops, attached to the cylindrical wall interact via reconnections and
expanding Kelvin waves and quickly fill the
cross-section of the cylinder within a short vertical section. The second
process is the expansion of this turbulence towards the vortex-free
region(s) as propagating turbulent vortex front(s) \cite{turbfront}. The axial propagation
velocity of the turbulent front $\Vf$ can significantly exceed the value which the laminar
expansion velocity $V_{\rm lam}$ would have in the same conditions.

In the experiment of Fig.~\ref{expsetup}A we first create counterflow by rotating the sample in the
vortex-free state (or a state with a few rectilinear vortices in a central
cluster) and then inject seed vortices. One injection technique uses the Kelvin-Helmholtz
instability of the interface between the A and B phases of superfluid
$^3$He \cite{KHprl}. Another
method is to start at a low rotation velocity $\Omi$ with a few remanent vortices
connected to the cylindrical wall and to create the flow by rapidly
increasing rotation to $\Omf$ \cite{mult-prl}. A third useful injection technique is triggered by an absorption event of a thermal neutron
which deposits, depending on the counterflow velocity, one or a few vortex loops of size $\sim 0.1\,$mm close to the
cylindrical wall \cite{neutron-prl}. After injection the seed vortices interact and evolve, expanding
along the cylinder. Using NMR techniques (Fig.~\ref{expsetup}C) we measure
the number of vortex lines close to both ends of the sample tube. If the vortex
expansion follows the laminar scenario, then they do not reconnect,
their number is not changed, and the height of the counterflow peak in the NMR spectrum decreases only a
little. If a turbulent front develops, then the counterflow peak disappears completely
when the front passes through a pick-up coil. There are practically no
intermediate cases, which cannot be classified as laminar or
turbulent. An example of a measurement of the turbulent dynamics with boundary-attached seed vortices is shown in Fig.~\ref{slowformation}.

If the vortex-injection measurement is repeated a number of times at the same
conditions, we obtain the probability for the turbulent
response. The results for different injection methods and different flow
velocities are shown in Fig.~\ref{probabs}. It reveals a sharp transition from
laminar to turbulent dynamics with decreasing temperature from which we determine the onset values $T_{\rm on}$
and $\Realon$ when the probability is 1/2. In agreement with the
criterion \eqref{eq:crit}, we find that $\Realon \sim 1$. Note that in the
case of the AB interface instability, when a small bundle of vortex loops is
injected in bulk B-phase, $\Realon$ is velocity-independent, as 
discussed above for bulk superflow. When the cylinder wall is closely involved in the turbulent processes, then the transition acquires some velocity dependence. However overall, the transition region is narrow, with the ratio
of $\Real$ being around~2 at its boundaries, when measured with a given injection
technique.

Although the cylinder wall approaches an ideal solid surface, its influence on the dynamics cannot be neglected at low mutual friction $\alpha \ll 1$: the drive from the counterflow reaches its maximum close to the side wall and in addition the self-induced velocity has to be taken into account which arises from the curvature of the vortex end at the attachment point on the wall. Thus reconnections with the wall, loop formation, and annihilation become important processes since their influence extends ever deeper into the bulk volume when $\alpha \rightarrow 0$. In the next two sections we discuss the influence of the boundary on the transition to turbulence and the dependence of $\Realon$ on the strength of the flow perturbation.

\subsection{Single-vortex instability: A precursor to turbulence}
\begin{figure}
\centerline{\includegraphics[width=\linewidth]{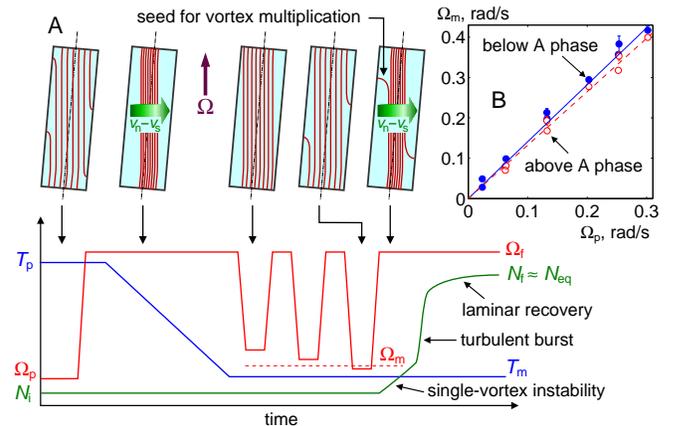}}
\caption{Instability in vortex -- wall interactions. (\textit{A}) Outline of the experimental
  procedures. The plot on the bottom shows the rotation velocity $\Omega$,
  the temperature $T$, and the number of vortices $N$ as a function of time, while
  the sketches in the top row illustrate the vortex configurations. See the text
  for details. (\textit{B}) Dependence of the rotation velocity
  $\Omega_{\rm m}$ at the instability on the preparation velocity $\Omega_{\rm p}$ (symbols) for the two
  B-phase sections, separated by the A phase barrier layer. The lower
  section is longer and thus the first vortex reconnects to the side wall
  at higher $\Omega_{\rm m}$. Both dependencies
  can be fit to a geometrical model of the vortex cluster \cite{tiltedcyl} using
  the residual tilt angle of the cylinder with respect to the rotation axis $\vartheta =
  0.64^\circ \pm 0.03^\circ$ as a fitting parameter (lines). The measurements here are
  performed with $T_{\rm p} = 0.75T_{\rm c}$, $T_{\rm m} = 0.4T_{\rm c}$
  and $\Omf - \Omega_{\rm p} = 0.7\,$rad/s.
\label{multseq}}
\end{figure}

The remarkable observation from the measurements in Fig.~\ref{probabs} is that the injection of even
a single vortex in the applied flow in a neutron absorption event results in a substantial probability to start
turbulence \cite{neutron-turbulence} already at a relatively small $\Real \approx 2$. What is the process 
responsible for the increase from one single vortex to a number
of loops which start interacting and produce turbulence? Insight in
this question is shed by the experiment outlined in Fig.~\ref{multseq}.
Here we start with the equilibrium vortex state at $\Omega = \Omega_{\rm
  p}$ where some vortices are curved and connect to the side wall owing to the residual tilt of the cylinder. At the initial high temperature $T_{\rm p} > T_{\rm on}$, rotation 
is increased to a sufficiently large value $\Omf$ so that all
vortices are collected in a  central cluster of rectilinear lines. The sample is then
cooled in this state to a temperature $T_{\rm m} < T_{\rm on}$. At this
temperature $\Omega$ can be increased or decreased, so that the vortex
cluster contracts or expands, and no turbulence is observed as long as no
vortices connect to the side wall of the cylinder. However, as soon as $\Omega$ is
reduced below the velocity $\Omega_{\rm m}$ where the first vortex reconnects
to the cylindrical wall, then on a subsequent increase of $\Omega$ to $\Omf$
turbulence evolves and ultimately brings $N \rightarrow N_{\rm eq}$.

The turbulent process consists of two stages. First during the
precursor, the number of
vortices $N$ increases linearly with time at a rate $\dot N \sim
1\,$s$^{-1}$ \cite{mult-prl,vorgen-jltp}. This slow increase is seen in Fig.~\ref{slowformation}. Later in the turbulent burst, $N$ suddenly increases close to the equilibrium number $N_{\rm eq}$. From these observations one concludes
that for starting the initial increase in $N$, it is necessary to have a
curved vortex, attached to the boundary and moving in the applied
flow. (In contrast, vortex ends on the flat top and bottom walls of the cylinder see approximately zero counterflow and are stable.)

\begin{figure}
\centerline{\includegraphics[width=\linewidth]{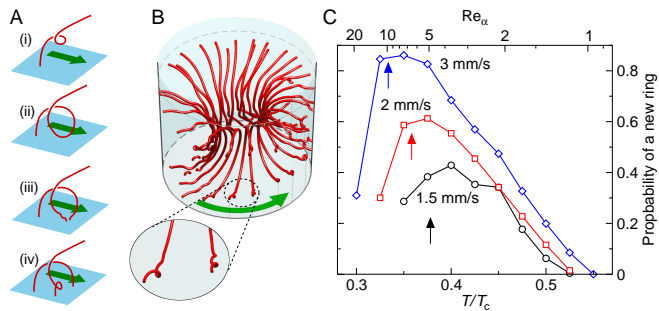}}
\caption{Instability of a curved wall-attached vortex in applied flow.
  (\textit{A}) Sketch of the sequence of events, which leads to the
  creation of a new vortex. See text for details. (\textit{B}) Snapshot from a numerical calculation 
  starting from a single vortex ring lying
  in the plane of the flow in a rotating cylinder. The snapshot is taken after the excitation of
  Kelvin waves on the ring and their reconnections with the
  cylinder wall. New Kelvin waves are seen to form on the wall-connected vortex segments. The later evolution creates a few hundred vortices via a
  series of single-vortex instabilities and the subsequent turbulent
  burst \cite{vorgen-jltp}. (\textit{C}) Probability to generate a new vortex  starting from a
  single ring with a radius of 0.5\,mm centered at a distance 1\,mm from
  the planar wall, along which counterflow is applied. The probability is averaged
  over all possible initial orientations of the ring. The counterflow velocity is
  marked at each curve. The arrows show the temperature where $\Real =
  \Relam/6$. By repeating the calculation with arbitrary combinations of
  $\alpha$ and $\alpha'$, it is proven that the result depends only on
  $\Real$, not on $\alpha$ or $\alpha'$ separately \cite{mult-prl}.
  \label{wallinstab}}
\end{figure}

A sketch of this flow-induced single-vortex instability at a solid surface
\cite{mult-prl} is
presented in Fig.~\ref{wallinstab}A. If a vortex close to the wall develops a
Kelvin-wave loop of proper orientation (i), the loop starts to grow in the
applied flow (ii). When the loop reconnects with the boundary, the reconnection
event creates a new vortex and induces Kelvin waves on the newly formed wall-attached vortex
segments (iii). In one of the segments the Kelvin waves will be properly
oriented with respect to the counterflow and the loop will start to grow (iv), repeating
the process. Thus new vortices are created repeatedly without the need for
a specially arranged 'vortex mill' \cite{Ruutu-vortform}. Eventually the local density of
vortices becomes sufficient to start the turbulent burst.

The main features of the above scenario are supported by numerical simulations,
although to reproduce the whole sequence of events in
Fig.~\ref{multseq} turned out to be difficult. In particular, in
Fig.~\ref{wallinstab}B one can see Kelvin waves on the side-wall-attached vortex
segments after the reconnection of a vortex loop with the boundary. In
Fig.~\ref{wallinstab}C a single vortex loop is allowed to expand and
reconnect with a flat wall. The probability to create a new vortex after
the reconnection is calculated, averaged over all possible orientations of the original
loop. Remarkably, this model demonstrates both central messages of
this paper. First, the probability to create a new vortex rapidly grows with
decreasing temperature in the range where $\Real \gtrsim 1$, demonstrating the
transition to turbulence. Second, at still lower temperatures, where $\Real$
approaches $\Relam$, the probability starts to decrease as a result of the
decoupling of the superfluid from the reference frame of the wall.

\subsection{Scaling properties of the transition to turbulence}
Above the transition to turbulence was studied by introducing a perturbation in metastable laminar flow and then observing
whether turbulence emerged or not. Similar investigations have been 
performed in viscous turbulence for the flow in a circular pipe \cite{mullin-prl,hof-science}. 
Laminar pipe flow is linearly stable which means that a finite-size
perturbation is needed to turn it turbulent. Measurements of the amplitudes $\varepsilon$ needed for
such a perturbation lead to a scaling law $\varepsilon \propto {\rm Re}^{-1}$.
Owing to the quantized nature of vortices in superfluids one might expect that a perturbation of finite amplitude is always required to turn the flow turbulent and thus the interesting
question arises whether a scaling law, similar to that of viscous
turbulence, might apply. Qualitatively a dependence on the amplitude
of the perturbation can be checked by comparing
different injection methods \cite{LTinj} and the respective values of $\Realon$. For example, as seen in
Fig.~\ref{probabs}, injection of $\Ninj\sim10$-20 closely packed vortex loops using
the AB interface instability results in a higher probability to start
turbulence at a given temperature and flow velocity compared to the
injection of $\Ninj\sim1$-3 vortex
rings using neutron absorption. Next we describe systematic measurements of
$\Realon$ using a method, where the strength of the flow perturbation,
expressed as the number of injected vortices $\Ninj$, can
be continuously adjusted.

This is achieved by varying the initial velocity $\Omi$ in the measurement
with curved wall-attached vortices in Fig.~\ref{slowformation}. A
simple geometric model of vortex lines in the tilted cylinder \cite{tiltedcyl} gives the number of vortex ends attached to the cylindrical wall as $\Ninj = (8\Omi R h /\kappa) (1 - \beta/2\sqrt{\Omi})
\sin\vartheta$. Here $h$ is the height of the sample, $\vartheta$ is the
tilt angle with respect to the rotation axis and the term with $\beta$ takes
into account the width of the equilibrium vortex-free annulus at the outer cylindrical boundary. 
Numerical calculations of the equilibrium vortex state in the tilted
cylinder give a smaller number for $\Ninj$ owing to the
effects from vortex curvature. However, in our search for scaling laws we are
not concerned with absolute numbers, but only use the dependencies $\Ninj
\propto \Omi$ and $\Ninj \propto h$.

\begin{figure}
\centerline{\includegraphics[width=\linewidth]{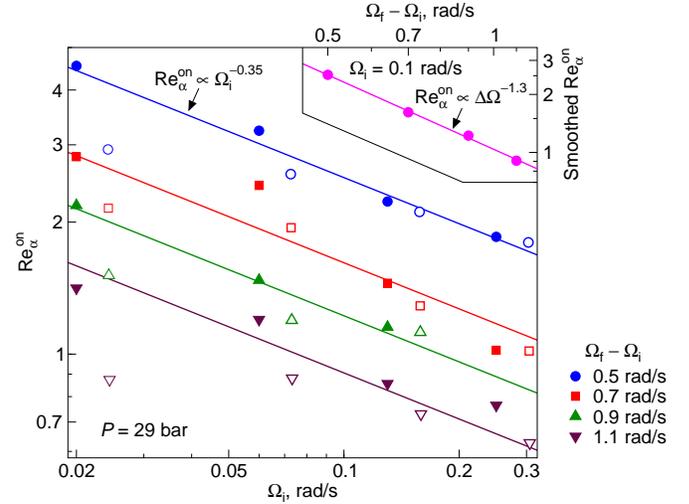}}
\caption{Scaling properties of the transition to turbulence with
  curved wall-attached vortices. The main plot shows $\Realon$
  as a function of $\Omi$, which controls the number of curved seed vortices
  at different values of applied flow $\Delta\Omega = \Omf - \Omi$. The filled symbols represent measurements on the
  shorter top B-phase section, while open symbols refer to the bottom
  section, see Fig.~\ref{expsetup}A. For the data on the bottom section, $\Omi$ has been scaled with $h_{\rm
    bot}/h_{\rm top}$, where $h_{\rm bot}$ and $h_{\rm top}$ are the
  heights of the two sections. The solid lines represent
  power-law fits to all data with a common exponent (excepting the bottom section at the lowest
  $\Omi$), but with a prefactor which depends on
  $\Delta\Omega$. The \textit{insert} shows $\Realon$ at $\Omi=0.1\,$rad/s
  as determined from the fits in the main panel. The line is the power-law
  fit for the data points.
\label{scaling}}
\end{figure}

Thus $\Ninj$ can be controlled by changing $\Omi$ and the applied flow velocity
with respect to the normal component, $\vs(R) = (\Omf - \Omi)R$, by varying 
$\Delta\Omega=\Omf-\Omi$. For each pair of $(\Omi,\Delta\Omega)$ we measure the 
probability curve \cite{PLTP}, as in Fig.~\ref{probabs}, independently for the two B-phase
sections and determine their respective onset values $\Realon$. The results are presented in
Fig.~\ref{scaling}. 

The transition is seen to move to higher temperatures
and smaller $\Real$ when either $\Omi$ or $\Delta\Omega$ is increased. In the
same conditions $\Realon$ in the longer bottom section of the sample container is
smaller that in the top section. The results from the two sections can be
put on the same line assuming proportionality of $\Ninj$ to $\Omi$ and
$h$. This supports the relevance of $\Ninj$ as a measure of the strength of the
perturbation. It is also understandable why the value of $\Realon$ in
the bottom section deviates downwards from the common
dependence at the lowest $\Omi$. This is caused by the rim of the orifice in the bottom
plate, where vortices become pinned and thus increase $\Ninj$.

Remarkably, the $\Realon(\Omi)$ dependence at all $\Delta\Omega$ can be fit with
the same power law $\Realon \propto \Omi^{-0.35}$ while the prefactor in
this fit also scales with $\Delta\Omega$ (insert in Fig.~\ref{scaling}).
When $\Realon$ is expressed in terms of $\Ninj$ using the conversion from $\Omi$ to
$\Ninj$ given above, the overall scaling becomes $\Realon \approx 2.2
\Ninj^{-0.3} \Delta\Omega^{-1.3}$. Note that in the previous section we
considered the instability of a curved wall-attached vortex in the applied
flow to develop independently of other vortices. The existence of the scaling proves that this view is oversimplified and the transition to turbulence is a collective phenomenon.
Beyond that, however, the understanding of the measured scaling is missing
and its explanation remains a task for future research. 

\subsection{Transition to turbulence in pipe flow}
The transition to turbulence in the flow through a circular pipe is one of the classic problems of hydrodynamics. In his pioneering work Osborne Reynolds \cite{Reynolds} demonstrated the influence of the
entrance to the pipe as well as the quality of his long straight flow channel on the downstream vorticity and turbulence. For superfluid pipe flow technical problems of this kind
have not yet been solved. However, numerical calculations can be performed to study the 
dynamics in superfluid pipe flow \cite{vorgen-jltp} and to check whether the principles emerging from rotating 
measurements also apply for linear flow in the pipe.

\begin{figure}
%\centerline{\includegraphics[width=0.8\linewidth]{pipeflow_example.eps}}
\centerline{\includegraphics[width=\linewidth]{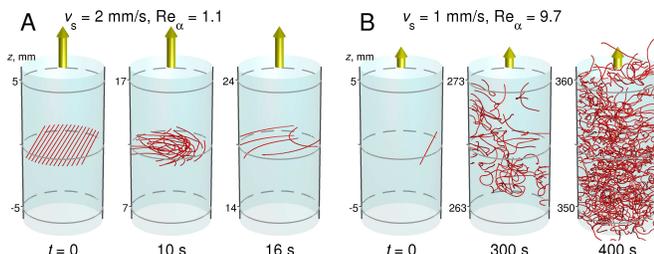}}
\caption{Two regimes of linear superfluid flow in a straight pipe of circular cross section. 
  These calculations assume $\vn = 0$ and a drive with a flat radial distribution of $\vs$ in the absence of vortices. The flow
  perturbation is introduced in the form of a few vortex lines stretched across the
  flow channel. The radius of the pipe is 3\,mm. (\textit{A}) At 
  small $\Real$, 15 vortex lines are
  unable to start turbulence and all vortices are eventually annihilated.
  (\textit{B}) At large $\Real$, one single vortex is sufficient to create
  a turbulent tangled vortex plug which moves downstream with the flow along the pipe. 
  \label{pipexmpl}}
\end{figure}

The first message is that turbulence is suppressed at high temperatures also
in pipe flow. Even when a number of vortex loops are introduced in the flow, so
that they start interacting and reconnecting, eventually they decay away by annihilating,
Fig.~\ref{pipexmpl}A. The reason is that under the influence of the mutual
friction force vortex lines move also transverse to the stream lines across the entire cross section of the pipe and annihilate on the wall, leaving no remnants. The second message is that at low temperatures
the single-vortex instability switches on and this radically 
changes the dynamics. Now the vortex, when it is driven to the wall across the pipe, does
not annihilate totally, but leaves two remnants, which start traversing the pipe
cross-section in the opposite direction. Such multiplying processes make it possible 
even for a single vortex to grow to a turbulent plug of tangled vortices,
Fig.~\ref{pipexmpl}B, analogous to the turbulent plugs of varying length observed in viscous pipe flow \cite{hof-science}. 

\begin{figure}
\centerline{\includegraphics[width=0.9\linewidth]{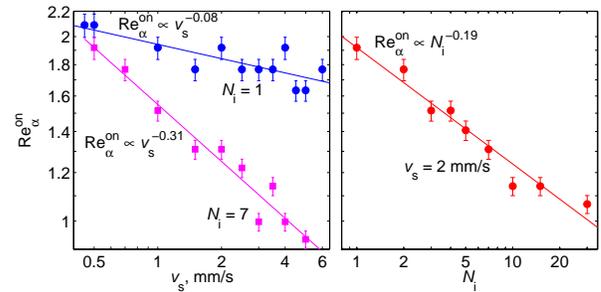}}
\caption{Scaling laws of the transition to turbulence for pipe flow from numerical calculations.
  The transition in terms of $\Realon$ is shown as a
  function of the flow velocity $\vs$ (left panel) and as a function
  of the number of seed vortex lines $\Ninj$ (right panel). The value of $\Realon$ is determined as the mid point between two calculations: one proving complete annihilation of all
  vortex lines at a smaller $\Real$ and a second demonstrating the formation of a
  turbulent vortex plug at a somewhat higher $\Real$. The separation of these two $\Real$ values is depicted with the uncertainty bars.
\label{pipescaling}}
\end{figure}

The transition separating the two regimes of superfluid pipe flow is sharp and occurs at $\Real \sim
1$, as demonstrated in Fig.~\ref{pipescaling}. The
onset value $\Realon$ scales with the
magnitude of the flow perturbation, expressed as the number of injected
vortices, in a manner similar to the measured dependence in rotating superfluid 
flow or in the pipe flow of classical liquids. Thus the calculations on 
superfluid pipe flow duplicate the essential features of the transition to turbulence, which were outlined
in the earlier sections.

\section{Energy dissipation and momentum transfer}

When the superfluid dynamics becomes turbulent at $\Real > \Realon$,
the assumption about the local polarization of vortex lines, used in
deriving equation \eqref{eq:vs}, starts to be violated. One way to
provide a phenomenological description of turbulent flows on the level of
the coarse-grained equations is to supplement \eqref{eq:vs} with effective friction
parameters. Effective friction models, though, do not possess universality: they depend
on the type of flow and on the particular physical
process considered. This is similar to how the scattering time of
quasiparticles in $^3$He is often treated: it is not a universal quantity, but depends eg. on
whether one considers viscosity, heat conduction, or mutual friction. So far the most detailed experimental information on  effective friction (or viscosity) was collected for turbulent energy dissipation
processes. Here we describe measurements which allow us to determine
simultaneously the effective frictions for energy dissipation and for
momentum transfer between the superfluid and its hydrodynamic drive \cite{FrontNatComm}. We
consider the lowest temperatures where $\alpha'$ can be ignored
and discuss only effective values of $\alpha$.

\subsection{Propagating turbulent vortex front} Fig.~\ref{expsetup}B
illustrates the emergence of an axially expanding process which ultimately
fills the rotating cylinder with rectilinear vortex lines in the
equilibrium state of rotation. The most exciting part of this
process is a turbulent vortex front \cite{turbfront} propagating along the entire length of the long cylinder. The front separates the vortex-free non-rotating superfluid from the rotating superfluid with a bundle of twisted vortex lines \cite{twisted-prl}, which we assume to be in approximate solid-body rotation. Thus $\vels$ is forced in a configuration of shear flow axially across the front. As usual for shear flow, such a configuration is unstable with respect to turbulence. At closer look it becomes obvious that vortices in the
front, which terminate on the cylindrical side wall, precess around the
cylinder axis at a smaller angular velocity than the twisted vortex bundle behind
the front. This differential motion leads to reconnections, which sustain the turbulence in stationary state while the front is propagating at constant velocity $\Vf$. 

The axial velocity $\Vf$ of the front is measured using NMR-based time-of-flight
techniques and is shown in Fig.~\ref{frontvel} as a function of temperature and rotation
velocity. One of the characteristics is that $\Vf$ saturates in the $T \rightarrow 0$ limit at a
$T$-independent, but $\Omega$-dependent value. Since the motion of the front
decreases the free energy of the superfluid, a non-zero $\Vf$ means that the
turbulence in the front leads to finite energy dissipation even
in the zero-temperature limit, in agreement with observations on the free
decay of vortex tangles in $^3$He-B \cite{Bradley}. Another non-trivial
feature of the front velocity is the temperature dependence
with a faster variation of $\Vf$ in the range $(0.2\div0.3)T_{\rm c}$, than
at higher or lower temperatures. This feature we call below the 'shoulder'.

\subsection{Decoupling of the superfluid from the container}
\begin{figure}
\centerline{\includegraphics[width=0.9\linewidth]{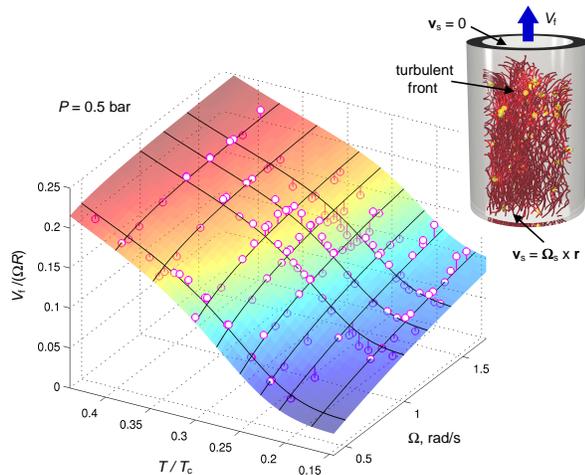}}
\caption{Propagation of the turbulent vortex front. The picture on the upper right is a snapshot from numerical
calculations of vortex expansion in the rotating cylinder at $ 0.27 \, T_{\rm c}$. In the region of the front the vortices terminate at the cylindrical side
wall. Turbulence is sustained by vortex reconnections which are highlighted with
yellow dots. The front moves upward into the vortex-free region. The plot on the left
shows measurements of the axial front velocity (circles) as a function of
temperature and angular velocity $\Omega$ of the cylinder. The lines are
fit to the model Eq.~\eqref{eq:model} with parameters $\Cen=0.52$, $\Cam=1.33$, $\taen=0.20$, and $\taam=0.0019$.
\label{frontvel}}
\end{figure}

\begin{figure}
\centerline{\includegraphics[width=0.8\linewidth]{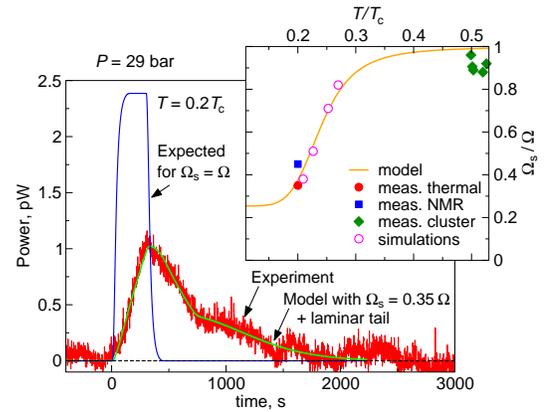}}
\caption{Decoupling of the superfluid from the reference frame of the
  container in the dynamics of the turbulent vortex front. The main panel
  shows the heat release during and after the propagation of the front. The
  front is triggered at $t=0$ and reaches the end of the sample cylinder at
  $t\approx 300\,$s, when the measured power is at the maximum. The
  square-wave model (blue curve) assumes the superfluid component to reach the equilibrium vortex state immediately behind
  the front. In fact, owing to its partial decoupling from the container rotation, the superfluid component comes to equilibrium much later, as the fitted analysis of the measured record shows (green curve). It proves that the vortex density behind the front
  corresponds to only $0.35\,\Omega$, while the rest of the vortices are
  supplied mostly by laminar motion well after the front propagation has
  already stopped. The insert shows the rotation velocity $\Oms$ of the
  superfluid behind the front as a function of temperature (normalized
  to the velocity $\Omega$ of the rotation drive). $\Oms$ is determined
  from an analysis of the thermal signal in the main plot, from oscillating
  NMR signals \cite{decouplingPRL}, and from direct measurements of the number of vortices
  behind the front at higher temperatures \cite{Finne-turbmeas-jltp} (filled symbols). Results from
  numerical simulations on front motion are shown with empty circles. The
  line represents the model in Eq.~\eqref{eq:model} with the same parameter
  values as in Fig.~\ref{frontvel} at $\Omega = 1\,$rad/s.
\label{decouple}}
\end{figure}

The shoulder in Fig.\;\ref{frontvel} falls in the temperature regime where
$\Real/\Relam \sim 1$ and the quasiparticle density approaches the
collisionless limit. Here momentum exchange of the superfluid component
with the normal excitations becomes increasingly weaker and the effects
from vortex tension and curvature gain importance in Eq.~\eqref{eq:vs}.
This leads to partial decoupling of the superfluid from the reference frame of the container
\cite{decouplingPRL}. It means that behind the front the superfuid
component rotates at an angular velocity $\Oms < \Omega$, that is at a rate
slower than what solid-body rotation in equilibrium with the cylinder wall
requires. In Fig.~\ref{decouple} measurements of the precession
velocity $\Oms$, as derived from the heat released in front propagation at
$T=0.2\,T_{\rm c}$, give a result which is less than half of that of the
rotation drive $\Omega$. This is consistent with $\Oms \approx 0.4 \, \Omega$
obtained from direct observations of vortex precession with NMR and from
numerical simulations at this temperature.

To account for the decoupling, one considers the angular momentum balance for the propagating front in Eq.~\eqref{eq:vs}, where $\fo{ns}$ appears since $\Oms \ne \Omega$ and $\fo{tens}$ owing to the vortex twist. In addition $\alpha$ has to be replaced by an effective value for
angular momentum $\aam$. For the energy balance, which in the laminar regime was expressed as $V_{\rm lam} \approx \alpha
\Omega R$, one then recovers a generalized expression with $\Omega$
replaced by $\Oms$ and $\alpha$ by a distinct effective value for the
energy dissipation $\aen$. The combined model is \cite{FrontNatComm}
\begin{equation}
\Vf = \aen \Oms R, \qquad
\Oms= \frac{\aam \Omega^2}  {\aam\Omega+\lambda R^{-2}}.
\label{eq:model}
\end{equation}
We assume a simple linear relation of the effective friction parameters with
$\alpha(T)$: $\aen (T) = \Cen\alpha (T) + \taen$ and $\aam (T) = \Cam\alpha (T) + \taam$,  with
$(T,\Omega)$-independent phenomenological parameters $\Cen$, $\taen$, $\Cam$
and $\taam$. This model provides a good fit to the experiment
in Fig.~\ref{frontvel} and reproduces the main qualitative features of $\Vf$
including the shoulder. The same parameter values also describe
measurements of $\Vf$ at different liquid pressures \cite{FrontNatComm} and
the direct
measurements of $\Oms$ in Fig.~\ref{decouple}. The systematic deviation of
the model from the experimental data in the range (0.25 --
0.35)$\,T_\mathrm{c}$ in Fig.~\ref{frontvel} may be caused by an
oversimplified treatment of the angular momentum balance in the turbulent
front which does not yet include all contributions considered for the
laminar front \cite{sonin-front}. 

An important message is conveyed by the residual values for $\taen=0.20$ and
$\taam=0.002$, which describe the $T\rightarrow 0$ contribution to
effective friction from turbulence and possibly from surface friction.  When energy dissipation is considered,
turbulence dramatically enhances effective friction, owing to the
turbulent energy cascade, which leads to substantial dissipation even in
the zero-temperature limit. The minute 
value of $\taam$, which is two orders of magnitude smaller than $\taen$, attests that
quantum turbulence does not efficiently increase momentum exchange with the normal
component and the container. As a result, a new class of hydrodynamic phenomena becomes prominent owing to superfluid decoupling at temperatures where $\alpha \ll 1 $ and $\Real \gtrsim \Relam \gg 1$.

\begin{figure}
\centerline{\includegraphics[width=0.85\linewidth]{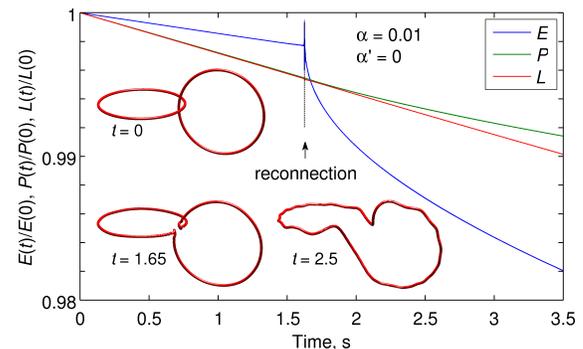}}
\caption{Numerical calculation of two reconnecting vortex rings.
  Two rings with a radius of 1\,mm are placed initially at a distance of 1.9\,mm
  between their centers, with their planes perpendicular to each other. The
  plot shows the calculated time dependences of the total kinetic energy $E(t)$, momentum $P(t)$,
  and angular momentum $L(t)$, normalized to their respective values
  at $t=0$. The calculation demonstrates quite different behavior in energy and momentum loss
  after the reconnection event. No externally applied flow is present.
\label{recrings}}
\end{figure}

The difference in the contribution from quantum turbulence to energy
dissipation and to momentum exchange with the normal component may
originate in the damping properties of the Kelvin waves, emitted from
vortex reconnections.  This is illustrated numerically in a simple
model system of two vortex rings, reconnecting in the presence of mutual
friction damping \cite{risto-rings}, Fig.~\ref{recrings}.  The damping causes a continuous
decrease of the total energy and momentum of the system. After the
reconnection Kelvin waves are excited and the oscillating motion causes
substantial extra energy dissipation, while the rate of momentum change is
essentially not affected by the reconnection event.

The explanation is that the direction of the mutual friction force alternates
in the two half-periods of the oscillation. Thus
momentum transfer to the normal component cancels out. Simultaneously the
direction of the vortex velocity is also alternating, but it remains 
opposite to the mutual friction force. Thus
the work done by
mutual friction in the two half-periods of the oscillation adds up. Since reconnections and the fluctuating
vortex motion are the characteristic features of quantum-turbulent flows,
this conclusion is likely to apply to quantum turbulence in general.

\section{Discussion}

The normal component can be considered not to participate
in the motion if $\nu \gg \kappa$ in the hydrodynamic regime while in the ballistic regime the
diffusive scattering of thermal quasiparticles from the walls should dominate the
scattering from vortices. These conditions are generally realized in superfluid $^3$He but may also be applicable to
superfluid $^4$He in the low-temperature range. When the normal component
is stationary and surface pinning or friction can be ignored, the motion
of the superfluid component is characterized by three dimensionless
parameters: $\Relam$, $\Real$ and their ratio.

The parameter $\Relam$ describes the transition from vortex-free or
single-vortex dynamics at $\Relam \lesssim 1$ to collective dynamics with
many vortices at $\Relam\gg 1$. The latter condition is required for
quantum turbulence. The parameter $\Real$ has the same physical meaning as the Reynolds number
in classical turbulence. When mutual friction is large and $\Real \lesssim
1$, turbulent motion is damped. One might think that by placing a
sufficiently dense vortex tangle in high-velocity flow, one could observe turbulent motion even at high
temperatures. In reality when $\alpha$ is large, vortex lines move across the
flow, i.e. towards the wall. If they annihilate at the wall without leaving
remnants behind, which is the case at an ideal wall and small $\Real$, then all
vortices eventually disappear. %(We do not consider here systems with effectively infinite initial vortex number or with the infinite extrinsic vortex supply.)

With decreasing temperature, i.e.\ increasing $\Real$, vortices start to move
with the flow, which allows more time for interactions and
reconnections. In this regime also the single-vortex instability at the wall \cite{mult-prl} becomes an
effective mechanism for the formation of new loops. Experimentally one can distinguish between the formation of new vortices in the bulk-volume or in surface-mediated processes since the latter switch on at
slightly lower temperatures. But overall both types of processes contribute to the
transition from laminar to turbulent superfluid dynamics at $\Real \gtrsim 1$.

The value of $\Real$ at the transition depends on the type of
flow and scales with the magnitude of the flow perturbation. In some
cases laminar flow can be exceptionally stable, like in the
spin-down of a cylindrical container to rest. Here a large perturbation, like tilting the
sample cylinder by $30^\circ$ from the rotation axis, is needed to turn the flow turbulent even at
$\Real \sim 10^3$ \cite{LaminarDecay}. This is in contrast to classical fluids,
where the spin-down to rest is one of the most unstable flows \cite{spindown-class}. The
difference is caused by the different mechanisms of coupling of the fluid
to the container: In classical fluids it is a surface effect, mediated by
boundary layers, while in superfluids it is a volume effect (in the absence of surface pinning or friction), mediated
by the normal component.

The implications from this coupling in low-temperature superfluid dynamics have
been understood only recently. It is by now well established that quantum
turbulence turns a superfluid to a dissipative liquid, by providing
intrinsic mechanisms for energy loss via the turbulent
energy cascade. In contrast, no such intrinsic mechanism exists for
momentum exchange with the normal component/container/drive, if extrinsic mechanisms such as surface friction
and pinning are excluded. With respect to momentum coupling and drag, a
superfluid filled with a vortex tangle continues to behave like an almost
ideal liquid and no boundary layer, similar to that in classical turbulence, is formed.

This duality in coupling mechanisms has profound implications on the
low-temperature dynamics, when $\Real/\Relam \gtrsim 1$. In
particular, the rapidly decreasing momentum coupling leaves a characteristic
anomaly in the overall energy dissipation rate, as demonstrated by the
'shoulder' in the measured front velocity $\Vf /(\Omega R)$ in
Fig.~\ref{frontvel}.  The anomaly falls in the temperature range where
another interesting phenomenon is expected to become important, the crossover in the turbulent energy
cascade from the quasi-classical cascade at length scales exceeding the
inter-vortex distance to the Kelvin-wave cascade at smaller scales. The problem of joining
the two energy cascades of different nature across the crossover region
\cite{lvov-bot,svistunov-bot,sonin-bot}, dubbed 'the bottleneck problem',
has still no universally accepted theoretical resolution. 
The
mutual-friction-determined termination point of the energy cascade traverses the
crossover region in the temperature range where $\Real^2/\Relam \sim 1$
\cite{volovik-regimes} or $\alpha \sim 10^{-2}$ \cite{lvov-suppression}.
This is not very different from the range where decoupling becomes
important owing to the rapid variation of $\alpha (T)$. It is thus not
trivial to distinguish the two effects experimentally. In fact, the
'shoulder' in the front velocity was originally interpereted as an indication of the
bottleneck \cite{turbfront}, and only newer direct measurements of $\Oms$
allowed to understand that decoupling is the main source of this anomaly.
An open question is whether the systematic deviation of the front
velocity from the decoupling model in the range (0.25 --
0.35)$\,T_\mathrm{c}$ in Fig.~\ref{frontvel} may provide new information on
the bottleneck.

A further consequence from the decreasing momentum coupling at low temperatures
is the increased importance of laminar flow in the low-friction
region, where all dynamics was earlier expected to be turbulent.
For example in the
thermal measurement of the front propagation in Fig.~\ref{decouple} only
about 40\% of the energy is released by the turbulent process and the rest
comes as slow laminar tail.
The reason is that in laminar flow the decoupling problem does not appear,
as both energy dissipation and momentum transfer are determined by the
same mutual friction parameter $\alpha$. 

An interesting question is to what extent the decoupling phenomena are
applicable to quantum turbulence in superfluid $^4$He, where usually
complicated experimental geometries and the pinning of vortices at the
sample boundaries facilitate the coupling of the superfluid to the
container. Simultaneously, however, much bigger values of $\Real$ can be
achieved in $^4$He compared to $^3$He which in principle increases the role
of decoupling. Suppression of the decoupling by pinning might be not
complete for the following reasons: First, the pinning force is limited by
vortex tension $T_{\rm v}$. This limit has been experimentally
observed in spin-up of $^4$He in rough-wall containers
\cite{pinning-spinup}. Second, it has been demonstrated that a vortex in
$^4$He can slide relatively easily along the wall by continuously
reconnecting from one pinned vortex to another
\cite{zieve-sliding-vortex}. Thus it may be worth to consider whether
the decoupling affects the interpretation of experiments in superfluid
$^4$He. One example is the measurements of spin-down of a cubic container
filled with superfluid $^4$He \cite{golov-prl}. Here the anomaly in the temperature
dependence of the effective turbulent viscosity $\nu^\prime$ was observed,
in a way similar to
the 'shoulder' in the front velocity $\Vf$ in our experiments. The current
explanation of this effect is linked
to the physics of the crossover region in the turbulent energy cascades
\cite{svistunov-scanning}. We note, however, that the decrease of
$\nu^\prime$ is observed in the temperature region where $\Real \sim \Relam
\sim 10^3$ and thus the influence of the decoupling can not be a priori
excluded. 

The superfluid decoupling is most pronounced in flows where the
momentum transfer from the drive to the superfluid is unidirectional. The opposite extreme is quantum turbulence generated with various oscillating objects \cite{osc-review}, currently the most popular technique for studying the free decay of turbulent vortex tangles \cite{Bradley}. Here the direction
of momentum transfer is continuously alternating, with the average being
close to zero. In this case the decoupling phenomena are masked. The recently
developed 'floppy' vibrating wire \cite{floppy}, which can be used in an
oscillating mode or in a steady-motion mode, might provide a new tool to study
the decoupling phenomenon, provided that surface-friction effects can be kept
under control.

\begin{acknowledgments}
We thank G.E. Volovik, V.S. L'vov and J.J. Hosio for useful discussions. 
This work was partially supported by the Academy of Finland (Centers of
Excellence Programme 2012-2017 and grant 218211) and the EU 7th Framework Programme (FP7/2007-2013, grant 228464 Microkelvin).
\end{acknowledgments}

\end{article}
\end{document}